\documentclass[]{aa}

\usepackage{natbib}
\usepackage{graphicx}
\begin{document}

\title{Truncation of galaxy dark matter halos \\ in high density environments}

\titlerunning{Truncated galaxy dark matter halos in clusters}
              
\author{M. Limousin \inst{1, 4}
  \and 
  J.P. Kneib \inst{2}
  \and
  S. Bardeau \inst{3,4}
  \and 
  P. Natarajan \inst{5, 6}
  \and
  O. Czoske \inst{7} 
  \and
  I. Smail \inst{8} 
  \and 
  H. Ebeling \inst{9} 
  \and  
  G.P. Smith \inst{10} 
}

\offprints{M. Limousin: marceau@dark-cosmology.dk}

\institute{
  Dark Cosmology Centre, Niels Bohr Institute, University of Copenhagen, Juliane Maries Vej 30, 2100 Copenhagen, Denmark
  \and 
  OAMP, Laboratoire d'Astrophysique de Marseille - UMR 6110 - Traverse du siphon, 13012 Marseille, France
  \and 
  L3AB - UMR 5804 - 2, rue de l'Observatoire, BP 89, 33270 Floirac, France
  \and
  Laboratoire d'Astrophysique de Toulouse-Tarbes, CNRS-UMR 5572 \& Universit\'e Paul Sabatier Toulouse III, 14 Avenue Edouard Belin, 31400 Toulouse, France
  \and
  Astronomy Department, Yale University, P.O. Box 208101, New Haven, CT 06520-8101, USA
  \and
  Department of Physics, Yale University, P.O. Box 208101, New Haven, CT 06520-8101, USA
  \and 
  Argelander-Institut f\"ur Astronomie, Universit\"at Bonn, Auf dem H\"ugel 71, 53121 Bonn, Germany
  \and 
  Institute for Computational Cosmology, Durham University, South Road, Durham DH1 3LE, UK
  \and 
  Institute for Astronomy, University of Hawaii, 2680 Woodlawn Drive, Honolulu, HI 96822, USA
  \and 
  School of Physics and Astronomy, University of Birmingham, Edgbaston, Birmingham, B15 2TT, England
}

\date{Received -- / Accepted --}


\abstract {} {Our aim is to constrain the properties of dark matter
halos inhabiting high density environments, such as is the case in massive 
galaxy clusters.} {We use
galaxy-galaxy lensing techniques that utilize a maximum likelihood
method to constrain the parameters of the lenses. It has been
demonstrated that such a technique provides strong constraints on the 
parameters that characterize a galaxy halo, as well as on the
aperture mass of these halos. In this analysis, we only use weak shear data and do not
include strong lensing constraints.} {We present the results 
of a study of galaxy-galaxy
lensing in a homogeneous sample of massive \textsc{x}-ray luminous clusters at
$z\sim$ 0.2. These have been observed in three bands with the \textsc{cfh12k}
instrument.  We find dark
matter halos in these clusters to be compact compared to those
inferred around isolated field galaxies of equivalent luminosity at
this redshift: the half mass radius is found to be smaller than 
50\,kpc, with a mean total mass of order $0.2\cdot10^{12}$ M$_{\sun}$.
This is in good agreement with previous
galaxy-galaxy lensing results and with numerical simulations, in particular with the
tidal stripping scenario. We thus
provide a strong confirmation of tidal truncation from a homogeneous
sample of galaxy clusters. Moreover, it is the first time that cluster
galaxies are probed successfully using galaxy-galaxy lensing techniques
from ground based data.}{}

\maketitle

\section{Introduction} 

Gravitational lensing has become a very powerful tool to measure the
mass distribution of structures in the Universe on a large range of
scales.  Both the strong lensing and the weak lensing regime are used
to map mass distributions. On large scales, the weak distortions
detected in the shapes of distant galaxies allows us to study the
distribution of matter on cosmic scales \citep{refregier03}.  On
galaxy cluster scales, the strong lensing features observed in the
cores of massive clusters can be used to put strong constraints on the
inner part of the cluster potential (\citealt{stronglensing};
\citealt{smith05}), whereas at larger cluster-centric radius, in the
weak lensing limit, the ellipticities of background galaxies give an
estimate of the shear field induced by the gravitational
potential of the foreground cluster.  On the scale of individual
galaxies, much work has been done as well on modeling and
understanding multiple quasar systems (\citealt{fassnacht};
\citealt{phillips}). These strong galaxy-galaxy lensing analyses
provide a test of structure formation in cold dark matter
models since lensing provides an interesting way to estimate the inner
slopes of density profiles, which can be compared to theoretical
expectations.

Weak galaxy-galaxy lensing studies provide constraints on the physical
parameters that characterize the dark matter halos of galaxies.  This
is accomplished directly using lensing, since the deformation in the
shapes of background galaxies produced by the foreground lenses
although weak is observationally detected statistically.
Galaxy-galaxy lensing has been used to get constraints on field
galaxies in different surveys (see Section 2 for a review). 
This effect has also been used
successfully to map substructure in massive galaxy clusters
(see Section 2.4), where most previous work has
utilized strong and weak lensing features in order to constrain the
properties of galaxy halos associated with the locations of bright
early-type galaxies in clusters. 

In this paper, we report the first detection of galaxy-galaxy lensing
in clusters from ground based observations \emph{without}
utilizing strong lensing. We demonstrate that this technique works
even without the constraints from multiple images but as a consequence
we need more foreground-background pairs. To illustrate this
we apply the technique to a well defined homogeneous sample of
clusters all at $z \sim 0.2$. Additionally, using 3 photometric bands, we apply
photometric redshift determination techniques to assign redshifts to
all the background objects. The details of our method have been
presented in a recent theoretical paper (\citet{limousin05}, hereafter
Paper I). It was demonstrated by analysing simulated data that
the technique works well and allows us to retrieve the characteristic
parameters that describe the dark matter halos and hence to put strong
constraints on the aperture mass of these halos.
We confirm the fact that galaxy halos in clusters are compact
compared to halos of field galaxies of equivalent luminosity.
This is a strong
confirmation since it relies on a homogeneous sample of galaxy
clusters, whereas most early work relied on a heterogeneous sample.
The results presented here are averaged on a cluster galaxy population
from the centre of the cluster up to $\sim$ 2 Mpc, which represent a
significant fraction of the virial radius (depending on the cluster: for the
sample we consider in this work, the virial radius spans from 2.5 Mpc for Abell~2218
to 3.6 Mpc for Abell~1835).

This paper is organized as follows: In Section 2 we review
galaxy-galaxy lensing results to date; Section 3 describes the data
and reduction techniques, as well as the procedure used to derive
catalogues from these images. The determination of photometric
redshifts is described in Section 4 and illustrated for the specific
case of Abell~1763.  The formalism used for modeling dark matter halos
and the maximum likelihood method are outlined in Section 5.  More
details on the analysis are given in Paper I. Our results are
presented in Section 6 , where we also make comparisons with other
galaxy-galaxy lensing results.  
Discussion of the results and conclusions are presented
in Section 7.  All our results in this paper are scaled to the flat,
low matter density $\Lambda$CDM cosmology with $\Omega_M = 0.3, \
\Omega_\Lambda = 0.7$ and a Hubble constant $H_0 = 65$
km\,s$^{-1}$ Mpc$^{-1}$. In such a cosmology, at $z=0.2$, $1''$
corresponds to $3.55$ kpc.

\section{Galaxy-galaxy lensing as a probe of galaxy properties}

In this section, we provide a review of galaxy-galaxy lensing
results to date: we begin by reviewing theoretical works and then
summarize observational results.

\subsection{Theoretical analyses}

Galaxy-galaxy lensing methods have been developed just after the first
detection of the phenomena (Brainerd, Blandford \& Smail 1996,
hereafter \textsc{bbs}). Different observational configurations were
considered: for field galaxies, the work and techniques have been pioneered by \citet{rix}
and in cluster by \citet{priya97}. Other theoretical works followed
by \citet{geigertheorie} and \citet{limousin05}, respectively in
cluster and both in cluster and in the field. \citet{priya97}
partitioned the mass of the cluster into a smooth clump and sub-halos
that are associated with early-type galaxies. Robust constraints are
obtained on both these components using strong and weak lensing
observations. The presence of a few multiply imaged systems in the
clusters that they modeled (with known measured redshifts) gave
tighter error bars on the properties of the dark halos that they
obtain. Other theoretical work does not include any strong lensing
constraints in the analysis.  The more recent study by
\citet{limousin05} went beyond the usual formulation and proposed a
re-parameterization of the problem in terms of more direct
physical quantities that allows putting strong constraints on the
aperture mass of a galaxy halo. These studies have demonstrated that
it is possible to recover the input parameters of the lenses. 
They used a generated
simulated catalogue defined to match present day observations in terms of
shape parameter measurements and object number density. It is found 
that the reliability of the galaxy-galaxy lensing signal depends on the number
density of galaxies whose distorted shapes can be reliably measured,
as well as any additional constraints that can be added to the
analysis, for instance, redshifts of the lens galaxies, redshifts of
the source galaxies, galaxy type, dynamical constraints, and the
presence of larger scale structure like groups or clusters in the
vicinity.

All these studies are based on a maximum likelihood analysis of the
data sets and the likelihood function is constructed from the
ellipticity probability distribution of galaxies.  Maximum likelihood
methods are preferred to the so-called direct averaging method (which
consists of obtaining an average shear field by simply binning up the
shear in radial bins from the centre of the lens outwards). The direct
averaging method is widely used to constrain galaxy cluster mass
profiles. In the case of galaxy-galaxy lensing, since the lensing
signal is much smaller than the characteristic noise which corresponds
to the width of the intrinsic ellipticity distribution of the
galaxies, one has to stack many individual galaxy shear profiles to
obtain a signal and to constrain an average galaxy halo population. Such
a method is possible when studying isolated field galaxies as
in \citet{fisher} study on the \textsc{sdss} data. The direct averaging method
supposes that we are able to isolate a lens in order to study it,
which is rarely the case. In fact, galaxy-galaxy lensing is fundamentally
a multiple deflection problem. This was first pointed out by \textsc{bbs} in their
early work, who found that more than 50\% of their source galaxies should 
have been lensed by two or more foreground galaxies: the closest lens on the sky
to any given source was not necessarily the only lens, nor the
strongest one. Moreover, \citet{brainerdconf} in an analysis of multiple deflections
by the galaxies in the Hubble Deep Field North (\textsc{hdf}) has shown that the probability of multiple
deflections exceeds 50\% for source galaxies with a redshift greater than 1
(see also the work by \citet{hoekstra05} where they consider isolated galaxies).
In many cases it has become clear
that it is almost never a unique lens that is responsible for the
detected lensing signal and that there are indeed no clean lines of
sight (\citealt{guzik}). Consequently, the problem 
is best tackled using an 'inverse'
method, and analysing galaxy-galaxy lensing using maximum likelihood
techniques is an example of such a method.

\subsection{Observational results}

The main goal of galaxy-galaxy lensing studies is to obtain
constraints on the physical parameters that characterize the dark
matter halos of galaxies. A dark matter halo can be described by two
parameters: in this work we will use $\sigma_0$, the central velocity dispersion,
which is related to the depth of the potential well, and
$r_{\mathrm{cut}}$, the cut off radius, which is related to the spatial
extension of the halo since it defines a change in the slope of the
three dimensional mass density profile: below $r_{\mathrm{cut}}$, the
profile falls with radius (see Paper I for a complete description of
galaxy dark matter halo modeling).
It should be noted that a dark matter halo parametrized by $r_{\mathrm{cut}}$
still has a significant amount of mass below $r_{\mathrm{cut}}$: the mass
profile become steeper, but half of the mass is contained below
$r_{\mathrm{cut}}$. Thus $r_{\mathrm{cut}}$ can be considered as a half mass radius.
More quantitatively, considering a galaxy sized dark matter halo with
$\sigma_0=220$ km\,s$^{-1}$ and $r_{\mathrm{cut}}=50$\,kpc, we derive
that $M(R>r_{\mathrm{cut}})$ = 45\% M$_{\mathrm{tot}}$.

In the following, we present galaxy-galaxy lensing results,
using $\sigma_0$ and $r_{\mathrm{cut}}$ to characterize their properties.
In practice, many galaxies have to be stacked in order to reliably detect a
signal. As the lenses considered in a given study do not have the
same luminosity, they cannot be assigned the same parameters. The standard approach is to use
scaling relations between the different lenses and to derive properties scaled
for a given luminosity L$^*$:
\begin{equation}
\sigma_0=\sigma_0^* (\frac{L}{L^*})^{\delta} \quad  \& \quad
r_{\mathrm{cut}}=r^*_{\mathrm{cut}} (\frac{L}{L^*})^{\alpha} 
\end{equation}

The case $\delta=0.25$ corresponds to the Faber-Jackson/Tully-Fisher relation.
The case $\alpha=0.5$ assumes that the mass to light ratio is constant for all
galaxies. These values are the one that are most often used in lensing
studies. However, there are other possible scaling relations, and
they can be tested with lensing.

We begin with the first \emph{non} detection of galaxy-galaxy lensing
by \citet{tyson}. Despite a vast amount of data (about 28\,000 
foreground-background pairs), they were unable to get a galaxy-galaxy
lensing signal, mainly because of the poor quality of the data at that time.
The first statistically significant detection of galaxy-galaxy lensing
is the work by \textsc{bbs} in 1996. They used deep ground-based imaging data
($\simeq$ 72 sq. arcmin.) to investigate the orientation of 511 faint
background galaxies relative to 439 brighter foreground field galaxies.
They claimed a detection of galaxy-galaxy lensing on angular scales
between $5\arcsec$ and $35\arcsec$ and derived limits on the characteristic
parameters of the dark matter halos of L$^*$ field galaxies : 
$\sigma_0=155 \pm 56$ km\,s$^{-1}$ and $r_{\mathrm{cut}} > 100h^{-1}$ kpc.
Since \textsc{bbs}, there have been different independent detections of galaxy-galaxy
lensing by field galaxies. We list them below and illustrate this enumeration on 
Fig.~\ref{comparfieldcluster} along with the constraints from
these studies on the ($\sigma_0,r_{\mathrm{cut}}$) plane (for the studies which
were able to constrain both parameters, with the values transferred to the cosmology
used in this paper). Studies on field galaxies are in grey.

\begin{itemize}
\item \citet{griffiths}, from the \textsc{hst} Medium Deep Survey, used 1\,600 foreground
objects ($15<\mathrm{\textsc{i}}<22$) and 14\,000 background objects ($22<\mathrm{\textsc{i}}<26$).
They were able to differentiate between spiral and elliptical lenses, and found:
$\sigma_0$ = 220 km\,s$^{-1}$ (elliptical) and 
$\sigma_0$ = 155 km\,s$^{-1}$ (spiral), as well as
a constraint on the halo extension : $r\simeq100 \, r_{\mathrm{hl}}$, 
where $r_{\mathrm{hl}}$ is the radius of the luminous component

\item \citet{dellantonio}, from \textsc{hdf} North, used a simple colour cut to differentiate
between lenses and sources: they used 110 lenses and 697 sources, and found
$\sigma_0 = 185^{+30}_{-35}$ km\,s$^{-1}$ and $r_{\mathrm{cut}}> 15\, h^{-1}$ kpc

\item \citet{hudson98}, from \textsc{hdf} North, with photometric redshifts for 208 lenses
and 697 sources, found $\sigma_0 = 148\pm28$ km\,s$^{-1}$ but no constraints on the
extension of these halos

\item \citet{ebbels}, from the \textsc{hst} Medium Deep Survey, used 22\,000 objects
and a magnitude cut, and found $\sigma_0 = 128^{+25}_{-34}$ km\,s$^{-1}$
and  $r_{\mathrm{cut}}>$ 120 kpc

\item \citet{fisher}, from the \textsc{sdss} data covering 225 sq. degrees, used
28\,000 bright objects ($16<r<18$) and 150\,000 background objects
($18<r<22$), and found $\sigma_0 = 145-195$ km\,s$^{-1}$
and $r_{\mathrm{cut}}> 275\, h^{-1}$ kpc
\item \citet{jaunsen}, from \textsc{cfrs} fields, with photometric redshifts,
found $\sigma_0 = 280\pm30$ km\,s$^{-1}$ and no constraints on the
extension of these halos

\item \citet{mckay}, from the \textsc{sdss} data with many more objects than the study by
\citet{fisher} and with spectroscopic redshifts for all lenses 
($3.4\cdot 10^4$ lenses with $r'<17.6$ and $3.6\cdot 10^6$ sources with $18<r'<22$),
found $\sigma_0 = $100-130 km\,s$^{-1}$ and $r_{\mathrm{cut}}>230\, h^{-1}$ kpc

\item \citet{smith01}, on the \textsc{lcrs}, used 790 lenses ($\mathrm{\textsc{r}}<18$) and
found  $\sigma_0 = 116 \pm 14$ km\,s$^{-1}$ and no constraints on the
extension of these halos

\item \citet{wilson}, studied elliptical galaxies in the redshift range
$0.25<z<0.75$. They used 15\,000 lenses with photometric redshift and 
148\,000 sources ($\mathrm{\textsc{i}}>25$).
No evolution in the velocity dispersion with redshift was established.
They found $\sigma_0 = 168^{+19}_{-21}$ km\,s$^{-1}$ and no constraints on the
extension of these halos

\item \citet{hoekstracnoc2}, from \textsc{cnoc-2} fields, with a magnitude cut
($17.5<\mathrm{\textsc{r}}_{\mathrm{bright}}<23$ and $22<\mathrm{\textsc{r}}_{\mathrm{faint}}<26$) found
$\sigma_0 = 133^{+14}_{-15}$ km\,s$^{-1}$ and $r_{\mathrm{cut}}= 260^{+124}_{-73}$ kpc
\item \citet{martinacombo17}, applied a maximum likelihood analysis on 
the \textsc{combo 17} survey (Classifying Objects by Medium-Band Observations in 17 filters)
where an accurate estimation of photometric redshift was possible. 
Considering all lenses, they found $\sigma_0 = 156^{+18}_{-24}$ km\,s$^{-1}$,
and $r_{\mathrm{vir}} = 209^{+24}_{-32}h^{-1}$kpc (1$\sigma$ confidence level).
Splitting the lens sample into two subsamples according to the spectral
types they found a 2$\sigma$ difference in the velocity dispersion which 
is larger for early-type galaxies.
Moreover, this work provides some constraints on the exponent of the scaling
relation on the velocity dispersion (see Eq.~1), finding 
$\delta=0.28^{+0.15}_{-0.12}$ for red galaxies, in agreement with
the Faber-Jackson relation.
\item \citet{hoekstra04} used Red Sequence Cluster Survey data (\textsc{rcs}).
Lenses were selected as objects as $19.5<\mathrm{\textsc{r}}<21$ and background
objects as $21.5<\mathrm{\textsc{r}}<24$. They found $\sigma_0 = 137 \pm 5$ km\,s$^{-1}$
and $r_{\mathrm{cut}}=185^{+30}_{-28}\, h^{-1}$ kpc
\item \citet{heymans06} from \textsc{hst gems} data. They constrain the evolution
of the virial to stellar mass ratio of galaxies with high stellar mass in the
redshift range $0.2\leq z \leq0.8$. The estimation of the stellar mass comes from
\textsc{combo 17}, and the measurements of the virial mass from a galaxy-galaxy
lensing analysis. They selected lenses by imposing a high stellar mass cut, and their
sample contains a majority of early type galaxies. Space observations provide
a number density of 65 source galaxies per square arcminute. They find a virial
radius of $\sim204\, h^{-1}$ kpc and a mean mass to light ratio
of $\sim123\, h$, in agreement with the findings of \citet{hoekstra05}.
These values can be compared to the one found in the presented work for
cluster early type galaxies (Table~1), attesting for a strong influence of the
environment on the galaxy properties

\item \citet{rachel06} from the \textsc{sdss}, present constraints on the halo
mass of the central galaxy and the fraction of galaxies that are satellites
as a function of $r$ band luminosity and stellar mass. Galaxy-galaxy lensing
was used to derive virial halo mass, and spectroscopy to derive stellar masses.
They looked at the efficiency with which baryons in the halo of the central galaxy
have been converted into stars, finding a factor of 2 or more difference in conversion
efficiency between typical spirals and ellipticals above stellar mass of
10$^{11}$ M$_{\sun}$.
They compare some properties of early type galaxies in both low and 
high-density region, and find both populations to have consistent central halo masses.

\end{itemize}

\subsection{Discussion}
Examining the studies above, we find that there is considerable
variation between data sets and the analysis techniques used by the
various authors. The imaging quality, size of the field, and the dichotomy
between lenses and sources differ significantly amongst these
investigations.  Most of them were limited to imaging in a single
bandpass hence they used a crude lens/source separation based upon
apparent magnitude.
Different groups consider different luminosities, make measurements
on various scales and therefore an exact comparison between each result
is difficult.
Moreover, the data are a heterogeneous mix of
deep images which were acquired for purposes other than galaxy-galaxy
lensing studies. 

Despite these differences one
has to keep in mind, the implications of these studies for the
physical characteristics of the halos of field galaxies are all
broadly consistent with one another,
which is remarkable, and which is the only robust conclusion that we want to draw
about the enumeration of the different detections on field galaxies.
Of course, we do not expect all the results to converge to a single set
of parameters because each study is different by itself as mentioned
before, and the interpretation of the characteristic
parameters depends on the morphological type of the galaxy hosted,
when most of the study were not able to split their lens sample between 
early and late type galaxies.

All the different studies do fit reasonable central
velocity dispersions, i.e. they are consistent
with results inferred from more traditional techniques such as
rotation curves.  In the case of galaxy halos in the
field no clear edge is detected to the mass distribution even on
scales of the order of a few hundred kpc (\citealt{mckay};
\citealt{fisher}). Only two published studies to date by
\citet{hoekstracnoc2} and \citet{hoekstra04} have been able to put an
upper bound on the characteristic extension of a field halo at about
$290^{+139}_{-82}\, h^{-1}$ kpc and $185^{+30}_{-28}\, h^{-1}$ kpc.
Besides these large values do not impose a stringent constraint for
typical galaxy mass distributions since at these radii the galaxy density
is only a few times above the mean density of the Universe.

These results on field galaxies are in rather good agreement with
studies based on satellite dynamics 
(\citealt{zaritsky93}; 1997; \citealt{prada}; \citealt{brainerd04}) 
where the idea is to use a
satellite galaxy as a test particle to probe the gravitational
potential of a brighter host galaxy that is considered to be more
massive. This method is feasible for isolated field galaxies, and the
authors have found the extension of dark matter halos of isolated
galaxies to be larger than 200 kpc.
The samples of host and satellites in current redshift surveys are
becoming large enough to be used to study the dark matter halos of 
the host galaxies: this is emerging as a powerful technique
that is entirely complementary to galaxy-galaxy lensing.

We can say that the last ten years since the first detection of
galaxy-galaxy lensing have been 'experimental' in the sense that these
early studies have demonstrated convincingly that galaxy-galaxy
lensing, though challenging to detect, is a viable technique by which
the dark matter distribution on scales of individual galaxies can be
investigated. Now that the technique has been proved, galaxy-galaxy
lensing shows great promise on getting interesting
statistical constraints on galaxy physics. At present the constraints 
obtained are not very strong, but the preliminary results are very 
encouraging. In particular, the following investigations are currently
pursued:
\begin{itemize}
\item halo parameter determination, mass measurement and the \textsc{m/l} ratio: 
evolution with redshift and influence of the
local environment, any evolutionary effects are investigated 
\item relation between the baryonic and dark matter components to provide constraints
for models of galaxy formation
\item deviation from spherical symmetry: there are both observational and 
theoretical arguments in favor of flattened halos, and galaxy-galaxy lensing
can provide constraints on the mean flattening of the dark matter halos of field
galaxies (see \emph{e.g.} \citealt{brainerd01}). \citet{hoekstra04} 
presented a weak lensing detection of the flattening of galaxy dark
matter halos, with an ellipticity $\sim 0.2$, implying that the halos are aligned
with the light distribution. Recently, \citet{rachel} detected an ellipticity from the 
\textsc{sdss} data set, which appears to be mildly inconsistent with the detection reported by
\citet{hoekstra04}. Note however that these two works used different data and
methodology, making a direct comparison difficult. 
Moreover, it is worth mentioning
that probing the shape of dark matter halos is still
a difficult measurement because one tries to measure an azimuthal variation of the
galaxy-galaxy lensing signal which is itself challenging to detect
\item morphological dependence of the halo potential:
dynamical studies suggest that
the depth of the potential wells of early-type $\mathrm{\textsc{l}}^*$ galaxies is deeper 
than those
of late-type $\mathrm{\textsc{l}}^*$ galaxies. The fact that early-type galaxies 
are more often observed
as acting as strong lenses than late-type galaxies \citep{stronglenses} reinforces this idea.
Some studies have been able to differentiate between spiral and elliptical
lenses (\citealt{griffiths}; \citealt{mckay}; \citealt{guzik}; \citealt{martinacombo17};
\citealt{hoekstra05}), and all
these studies have found their elliptical galaxy sample to produce stronger
galaxy-galaxy lensing signal than their spiral galaxy sample. \citet{mckay}
converted their galaxy-galaxy lensing signal to an aperture mass in an aperture radius
of $260\, h^{-1}$kpc and found it to be a factor 2.7 larger for ellipticals than for spirals.
\citet{guzik} in their work on the \textsc{sdss} data, also found that the virial mass \textsc{m}$^*$
of an \textsc{l}$^*$
galaxy varies significantly with galaxy morphology, with \textsc{m}$^*$ being lower
for late types relative to early types (up to a factor 10 in the $u'$ band).
It is interesting to find a similar result by using satellite dynamics: \citet{brainerd04}
computed the velocity dispersion profile for the satellites of host galaxies in the Two
Degree Field Galaxy Redshift Survey and in the $\Lambda$\textsc{cdm gif} simulation. She found the
velocity dispersion profile to have a substantially higher amplitude and steeper slope
for satellite of early-type hosts than it does for satellites of late-type hosts
\item the scaling of the total galaxy mass with luminosity, including any strong
evolution of these relations with redshift: \citet{mckay} split their lens
sample with luminosity, and found the shear signal to be strongly dependent on the luminosity:
the more luminous the lenses, the stronger the shear they produce, hence the more 
massive they are. 
These scaling laws have been confirmed by an
independent dynamical method applied on the same data set \citep{mckay02}.
\citet{hoekstra04} also provide a constraint on 
the scaling relation between the \textsc{b}-band luminosity and the velocity dispersion, and
found a relation that is in very good agreement with the Tully-Fisher relation.
Moreover, \citet{hoekstra05} considered isolated galaxies from the \textsc{rcs}: they split their sample
into 7 luminosity bands and measure the mean tangential shear signal out to 2 arcminutes
from the lens. They find that the strength of the lensing signal increases with the luminosity 
of the lens. As a consequence, the virial mass is found to be an increasing function of the
luminosity, with a slope of $\sim$1.5 in \textsc{b}, \textsc{v} and \textsc{r} bands.

\item the truncation of the dark matter halos during the infall of galaxies into cluster (see below)
\item the bias of light compared to mass by studying the galaxy-mass correlation function \citep{seljak05}
\item the nature of the dark matter: \citet{priyaA2218} have shown that constraints on the
extent of the mass distribution around galaxies in the rich cluster Abell~2218 obtained
from combining strong and weak lensing observations are consistent with the predictions
which assume that the dominant mass component (dark matter) in these halos is collisionless.
A strongly interacting (fluid-like) dark matter is ruled out at a confidence level of more
than 5$\sigma$
\item comparison between the virial and the stellar mass and constraints on the star formation 
efficiency (\citealt{heymans06}; \citealt{hoekstra05}; \citealt{guzik}; \citealt{rachel06}).
\end{itemize}

Future prospects are very promising: in particular, the exploitation of the \textsc{cosmos}
survey is ongoing. This survey is a 2 sq. degree imaging survey with the 
\textsc{hst}; it will contain about 10$^6$ galaxies and about
35\,000 spectra of galaxies to be measured with the \textsc{vimos} instrument on the \textsc{vlt}.
Moreover, the multi wavelength observations will assign secure photometric
redshifts for all objects.
As demonstrated by \citet{martinaz}, knowledge of the lens redshifts is very important
in any galaxy-galaxy lensing study. Applying galaxy-galaxy lensing techniques
to the full \textsc{cosmos} data will give interesting results on galaxy physics, and
will relate the galaxy properties to their redshift and the local environment.
Furthermore, the next generation of space telescopes will
allow probing deeper in the Universe by combining wide field and very high quality
data from space.
From the ground, ongoing or future surveys are also very promising for
galaxy-galaxy lensing studies (the \textsc{cfhtls},
the second generation of \textsc{rcs} survey, as well as the KIlo Degree Survey (\textsc{kids})).

\subsection{Galaxy-galaxy lensing through clusters:}
\begin{figure}[h!]
\begin{center}
\includegraphics[height=8cm,width=8cm]{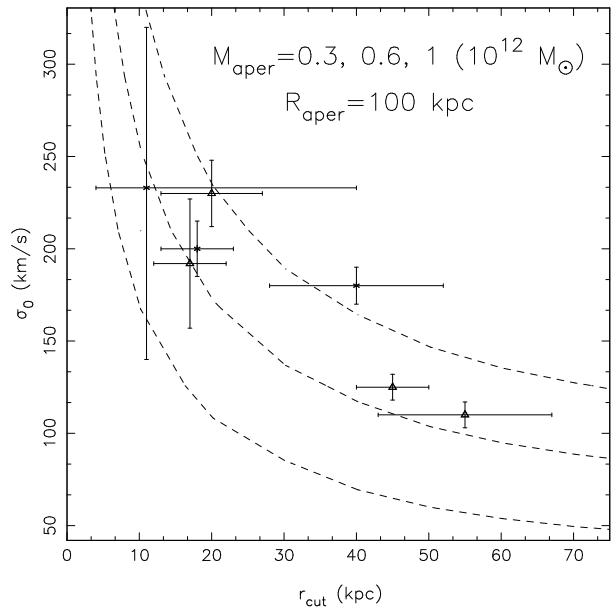}
\caption{Results obtained on cluster galaxies from galaxy-galaxy 
lensing analysis. Along the dotted lines, the mass within a 
projected radius $R_{\mathrm{aper}}\,=\,100$ kpc is
constant, equal to the value indicated on the plot}
\label{comparcluster}
\end{center}
\end{figure}

Galaxy-galaxy lensing has been used successfully
to map substructure in massive lensing clusters (\citealt{priyaAC114}; \citealt{geigeramas}; 
Natarajan et~al. 2002a, 2002b and the work presented in this paper).
Fig.~\ref{comparcluster} shows the different galaxy-galaxy lensing 
results on cluster galaxies.
Analyses on cluster galaxies all used \textsc{hst} data for their investigations.
The study of \citet{geigeramas} on galaxy cluster Cl0939+4713 led to
a detection of galaxy-galaxy lensing, but the field appeared to be too small
to allow strong conclusions to be drawn about the mass distributions of the
cluster galaxies.
Studies led by Natarajan et al. included strong constraints from observation
of multiple images systems which boost the convergence of the likelihood,
which is why their error bars are tighter than in the
study by Geiger \& Schneider. 

\begin{figure}[h!]
\begin{center}
\includegraphics[height=8cm,width=8cm]{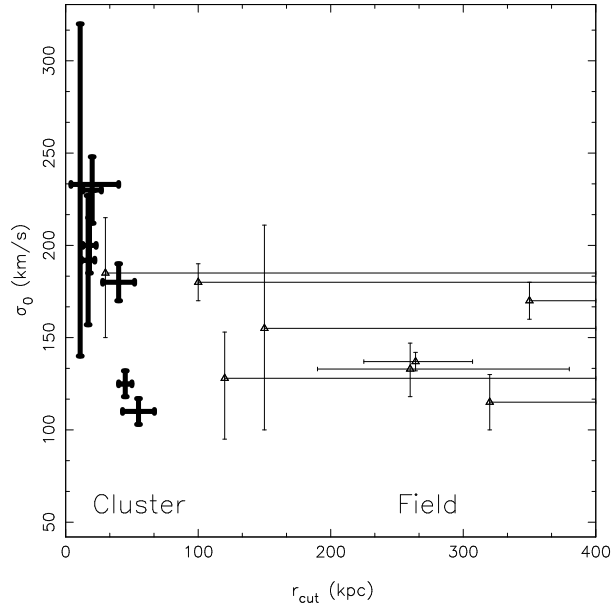}
\caption{Comparison between galaxy-galaxy lensing results on cluster galaxies
(black) and galaxy-galaxy lensing results on field galaxies (grey). 
References are given in section 2.2}
\label{comparfieldcluster}
\end{center}
\end{figure}

From Fig.~\ref{comparfieldcluster}, a clear trend can be seen: dark
matter halos in cluster are significantly more compact compared to
halos around field galaxies of equivalent luminosity.  This is a
landmark observational result from galaxy-galaxy lensing
studies, that was expected from theoretical considerations and
numerical simulations: when clustering, galaxies
experience strong tidal stripping from the cluster potential, and they
loose part of their dark matter halo, feeding the global cluster dark
matter halo itself.  Moreover, \citet{priyatidalstrip} have considered
five galaxy clusters which span a wide range of redshifts (0.18 $<z<$
0.58) and they find that not only are the dark matter halos truncated
in dense environment, the proper length of the truncation radius
increases with the redshift as expected from tidal stripping scenarios
\citep{ghigna98}.
However, given the small field of the \textsc{hst} data used there are likely to be
limitations from systematics.  The clusters used are a very
heterogeneous sample (they span a wide range in redshift, richness,
mass, \textsc{x}-ray luminosity) and possibly an increased proportion of
contaminating field galaxies for the higher redshift clusters.


\section{Observations and cataloging}
\subsection{Data}

The data used in the presented work were taken at the \textsc{cfht} with the 
\textsc{cfh12k} camera through the \textsc{b}, \textsc{r} and \textsc{i} filters.
A detailed description of the data acquisition and reduction can be found in
\citet{olliphd}. For a brief outline see \citet{bardeau05}.
No further details will be given in this paper.

The average seeing of the observations as estimated from the \textsc{fwhm}
of stars is 0.8 $\arcsec$ in the \textsc{r} band.
The average limiting magnitude in the \textsc{r} band is equal to 26.2

\subsection{From images to catalogues}

For a detailed description of the object detection see
\citet{bardeau05}.  Here we just give a brief outline of the different
steps involved in the analysis of the reduced and calibrated images.
A crucial step is to estimate the Point Spread Function (\textsc{psf}) and its
variation in each region of the images. The \textsc{psf} measures the response
of the entire optical system (atmosphere + telescope optics) to a
point source. In our case, stars provide
the calibrating point source, thus the shapes of stars detected in the
images provide our estimate of the \textsc{psf}. The shape of a star
includes an isotropic component mainly due to atmospheric seeing, as
well as an anisotropic component caused, for example, by small
irregularities in the telescope guiding.  The isotropic component of
the \textsc{psf} leads to a circularization of the images of small galaxies and
thus reduces the amplitude of the measured shear.  The anisotropic \textsc{psf}
component introduces a systematic component in galaxy ellipticities
and thus causes a spurious shear measurement if not corrected
\citep{kaiser95}.

The images were processed through various software routines in order
to extract the quantities we are interested in: the position of each galaxy,
the shape parameters and their magnitudes. The first step is to construct a
photometric catalogue for each individual image. In order to get the object
positions and magnitudes, we have used \textsc{sextractor}
\citep{Bertin-Arnouts1996}.  The second step of the analysis is to
extract a star catalogue from the full catalogue which will be used to
estimate the local \textsc{psf}.  We selected stars and
cleaned the resulting catalogue as described in \citet{bardeau05}. In order to
measure the shapes of the stars, we used the \textsc{im2shape} software
developed by \citet{bridle01}.  At this stage, we have a map of the
\textsc{psf} distribution over the entire field.  The third step is to compute the
galaxy catalogues that will be used in the weak lensing
analysis. Galaxies are selected from the photometric catalogues
according to the criterion described in \citet{bardeau05}.  To
measure the shapes of galaxies, we first linearly interpolate the
local \textsc{psf} at each galaxy position by averaging the shapes of the five
closest stars. This number of stars is found to be large enough to
locally interpolate the \textsc{psf}, whereas choosing a much larger number would
over-smooth the \textsc{psf} characteristics. \textsc{im2shape} then computes
the intrinsic shapes of galaxies by convolving a galaxy model with the
interpolated local \textsc{psf}, and determines which one is the most likely by
minimizing residuals. In the end, \textsc{im2shape}'s output gives a
most likely model for the fitted galaxy characterized by its position,
size, ellipticity and orientation, and errors on all of these quantities.

Finally, a master catalogue is produced which matches the objects
detected in the three filters and which contains colour indices built
from aperture magnitudes in 16 pixels (3.28$\arcsec$) diameter
apertures. This catalogue is used to plot the colour-magnitude
diagrams from which the sequence of elliptical is identified and
extracted.  \citet{bardeau05} used the full objects catalogues in
their weak lensing analysis.  In this paper, we use only objects
detected in all three bands and with reliable shape information.
Reliable shapes refers to objects for which the error on the
ellipticity is small.  If $e=e_1+i\,e_2$ is the complex ellipticity,
we impose the errors on these parameters to be err$(e_1)<0.1$ and
err$(e_2)<0.1$. Moreover, the shape parameters used in the final maximum
likelihood analysis are the ones derived from the \textsc{r} band, as the data quality
in terms of seeing and source density is superior compared to the other bands.
These objects
with three colours constitute the basis of our galaxy-galaxy lensing
analysis and we undertake a photometric study of these objects to
derive a redshift estimation for each object.

\section{Bayesian photometric redshifts}

Getting photometric redshifts with three bands is quite challenging
but possible and reliable for certain redshift ranges which are well
constrained by the filters we have. Adding a prior probability allows
us to get better constraints than we would have without any
assumptions. The method implemented here has been developed by
\citet{benitez}.  We will first verify the calibration of the
magnitude, and then show the results of a theoretical study we undertook in
order to quantify the kind of information we can derive from the data at hand.  
Then we will verify that the photometric redshift determination
is correct.

\subsection{Verification of the magnitude calibration}

Before using magnitudes in the analysis, we need to verify that they are
well calibrated.  From the colour-magnitude diagram, we are able to define
the \emph{most luminous} elliptical galaxies of the cluster on the red
cluster sequence (about 50
objects).  We then compare their colours to the Coleman, Wu \& Weedman
(\textsc{cww}) templates, which are  found to be a better comparison set to our
galaxies than the \citet{bruzual03} elliptical template. We selected
the most luminous objects because the \textsc{cww} template corresponds to a
metallicity close to solar metallicity.  Moreover, we choose the
\textsc{cww} template because it comes from observations of elliptical
galaxies in the local universe, and we can reasonably assume that
there is little evolution between the redshift of the cluster sample
($z\sim0.2$) and the present day.
The magnitude calibration appeared to be correct.

\subsection{Theoretical analysis}

\begin{figure}
\begin{center}
\includegraphics[height=4.35cm,width=4.35cm]{5543f3.1.ps}
\includegraphics[height=4.35cm,width=4.35cm]{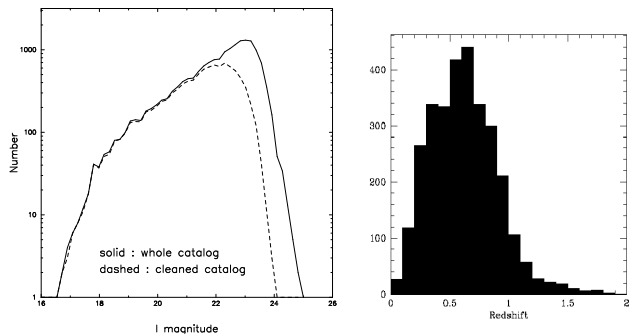}
\caption{Left: \textsc{i} magnitude distribution for the whole catalogue (solid line) 
and for the cleaned catalogue (dashed line). Right: redshift distribution 
from \textsc{vimos vlt} Deep Survey galaxies that verify \textsc{i} magnitude $<$ 24}
\label{Imag}
\end{center}
\end{figure}

Using the \textsc{hyperz} \citep{hyperz} templates, we are able to
simulate a catalogue with known $z_{\mathrm{model}}$ and the
corresponding \textsc{b}, \textsc{r} and \textsc{i} colours. These colours are then fed back to
\textsc{hyperz} in order to derive an estimate of the photometric
redshift $z_{\mathrm{phot}}$ (to compute the magnitudes, a noise is
estimated as a function of the apparent magnitude as explained in
\citet{hyperz}). 
We can delineate "good" and "bad" regions 
with respect to \emph{e.g} the criterion proposed by \citet{rix}: 
a region
is considered as a good one if the points do verify the following
constraint:
$0.5 < z_{\mathrm{phot}}/z_{\mathrm{model}} < 1.5$: 
\begin{itemize}
\item from $z=0$ to $z=0.5$, the constraints derived from \textsc{hyperz} are bad;
only 30\% of the objects satisfy the criterion
\item from $z=0.5$ to $z\sim 1$, we have a good region, with 60\% of the
objects that satisfy the criterion
\item from $z\sim 1$ to $z=3.7$, 73\% of the objects do satisfy the criterion
\item from $z\sim 3.7$ to $z=5$, 89\% of the objects do satisfy the criterion,
note that with increasing redshift, this criterion becomes less difficult to
satisfy.
\end{itemize}
We now restrict our study to the redshift range $z=0$ to $z=1.5$
because this is the range where we are likely to have the majority of
background objects:
\begin{itemize}
\item by comparison to the \textsc{hdf} \citep{bardeau05}, we found that the 
mean redshift of the background population is $z\sim1$
\item moreover, Fig.~\ref{Imag} shows the \textsc{i} magnitude distribution 
for the whole catalogue of Abell~1763 and for the cleaned catalogue, 
i.e. when we restrict ourselves to the objects
with a good measurement of the shape parameters.
We see that the magnitude limit for the clean catalogue is about 24.
Our catalogue can be compared to the \textsc{vimos} "deep field" \citep{vimos}, for which 
a redshift distribution is known: they use
a sample of galaxies $\mathrm{\textsc{i}}<24$ and from
the redshift distribution for this sample,we see (Fig.~\ref{Imag})
that most of the objects
are located at $z<1.4$, with a tail out to z=2 (5\% contamination above 1.4).
By comparison, we deduce that most of our objects are located at redshifts
lower than 1.4.
\end{itemize}

The second step for the analysis is to perform a Bayesian photometric study
for these simulated objects. The idea is to add a prior probability
which is not used by \textsc{hyperz} and in which we are confident.
This extra information will add constraints to get the redshift distribution of
an object. We will use as a prior the luminosity function.
The final redshift assigned to a galaxy is determined by
combining the information coming from the \textsc{hyperz} probability
distribution with the prior probability.  
Basically, adding this prior allows us to get rid of some degeneracies in 
the redshift probability distribution coming from \textsc{hyperz}:
using only three filters, 
the spectral energy distribution of a given galaxy is not well constrained,
thus the redshift probability distribution sometimes exhibits two solution:
a moderate and a high redshift one. The prior then ``kills'' some unlikely
high redshift solution, for example when a relatively bright galaxy has a 
redshift probability distribution with a moderate redshift solution ($z\sim0.5-1$)
and a high redshift solution ($z\sim3$).
We can delineate ``good'' and ``bad'' regions 
as before:
\begin{itemize}
\item from $z=0$ to $z=0.5$, only 40\% of the objects do satisfy the \citet{rix} criterion.
It is better than when not adding the prior, but still the redshift is poorly constrained in this
range
\item from $z=0.5$ to $z=1.5$, 98\% of the objects do satisfy the criterion.
\end{itemize}
We see that having a Bayesian approach to the problem by adding a prior 
probability does improve the redshift determination significantly in the range where it was 
already reliable before adding this prior.
Note however that this theoretical analysis is idealized in the sense
that it is based on synthetic galaxy templates and does not include any contamination by stars.
The redshift determination is quite reliable for the background population ($z>0.5$). 
These objects will be used as the background sample in the galaxy-galaxy lensing 
analysis.
On the other hand, the redshift of the lenses is not well constrained by our filters.
In particular, the cluster ellipticals are assigned a Bayesian redshift between
0.35 and 0.45, systematically overestimating their redshift.
So we decided to extract the elliptical galaxies from a colour magnitude diagram 
and assign them a redshift equal to the redshift of the cluster.
These objects will be the lenses in the following analysis. We are aware that we will
loose some lenses in doing so, but the advantage is to have the right redshift for
the cluster elliptical galaxies which are supposed to dominate the galaxy-galaxy
lensing signal.

\subsection{Reliability of the Bayesian photometric redshift determination: comparison
to the \textsc{deep2} survey}

\begin{figure}
\begin{center}
\includegraphics[height=8cm,width=8cm]{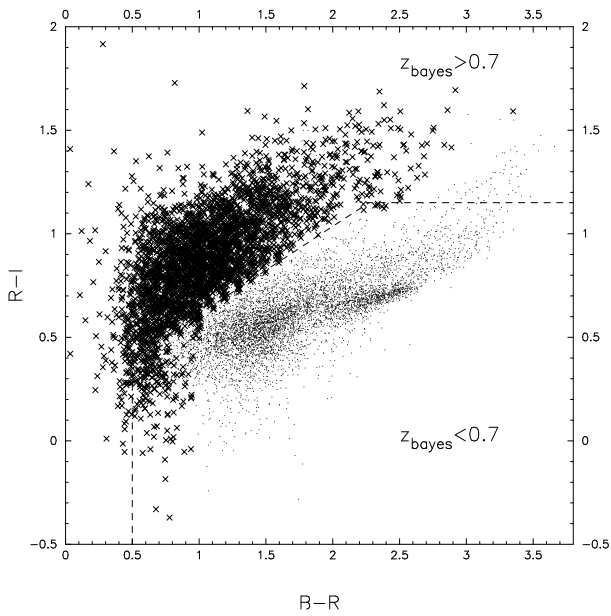}
\end{center}
\caption{Objects of our Abell~1763 catalogue with $z_{\mathrm{bayes}}>0.7$ (crosses)
and with $z_{\mathrm{bayes}}<0.7$ (dots).  The line 
represents the simple colour-cut used in the \textsc{deep2} survey to select 
objects with $z>0.7$ and $z<0.7$}
\label{deep2}
\end{figure}

To verify the reliability of the Bayesian photometric redshift estimation, we
compare with the \textsc{deep2} redshift survey \citep{deep2}. In the \textsc{deep2} galaxy
redshift survey, they use a simple colour-cut designed to select
galaxies at $z>0.7$: $\textsc{b}-\textsc{r} < 2.35 \times (\textsc{r}-\textsc{i}) - 0.45$, 
$\textsc{r}-\textsc{i} > 1.15$
or $\textsc{b}-\textsc{r} < 0.5$. 
As discussed in \citet{davis}, this colour-cut has proven effective: it
results in a sample with $\sim$ 90\% of the objects at $z>0.7$, missing
only $\sim$ 5\% of the $z>0.7$ galaxies.
We check to see where our objects with
$z_{\mathrm{bayes}}>0.7$ and $z_{\mathrm{bayes}}<0.7$ fall in a
colour-colour diagram, with respect to this colour-cut.  Figure
\ref{deep2} shows the results. The dashed line represents the
colour-cut: objects at $z>0.7$ are supposed to be above the regions
defined by these three dashed lines according to the \textsc{deep2} colour-cut.
The points represent our objects for which we have estimated
$z_{\mathrm{bayes}}<0.7$.  The crosses represent our objects for which
we have estimated $z_{\mathrm{bayes}}>0.7$. We can see from this plot
that our estimation of the redshift agrees well with the
colour-cut.
The next section gives an outline of the methodology used in this work.

\section{Methodology}

Full details on the methodology are given in Paper I. Here we briefly
explain the method: we introduce the parameters used to describe
the galaxy mass profile. Then we outline
the maximum likelihood analysis used to derive constraints on dark
matter halos of galaxies.

\subsection{Modeling the mass distribution of galaxies}

Beginning with the three dimensional density distribution $\rho(r)$
that fully characterizes a dark matter halo for our purposes, we project it onto the
lens plane to get the two dimensional potential, $\phi(R)$. The
related projected surface mass density $\Sigma$ is then given by:
\begin{equation}
\label{equ1}
4\pi\mathrm{G}\Sigma(R)=\nabla^2\phi(R)
\end{equation}
and the two-dimensional projected mass inside
radius $R$ (the aperture radius $R_{\mathrm{aper}}$) is defined as follows:
\begin{equation}
\label{equ2}
M_{\mathrm{aper}}(R)=2\pi\int_0^R\Sigma(r)rdr
\end{equation}

To model dark matter halos, we use 
the two components Pseudo-Isothermal Elliptical Mass Distribution 
(\textsc{piemd}, \citealt{kassiola}),
which is a more physically motivated mass profile than
the isothermal sphere profile (\textsc{sis}) but sharing the same profile slope
at intermediate radius.
The density distribution for this model is given by:

\begin{equation}
\label{rhoPIEMD}
\rho(r)=\frac{\rho_0}{(1+r^2/r_{\mathrm{core}}^2)(1+r^2/r_{\mathrm{cut}}^2)}
\end{equation}
with the core radius $r_{\mathrm{core}}$ of the order of $0.1\arcsec$, and a
truncation radius $r_{\mathrm{cut}}$. In the centre,
$\rho\sim\rho_0/(1+r^2/r_{\mathrm{core}}^2)$ which describes a core with
central density $\rho_0$. The transition region 
($r_{\mathrm{core}}<r<r_{\mathrm{cut}}$)
is isothermal, with $\rho\sim r^{-2}$. In the outer parts, the
density falls off as $\rho\sim r^{-4}$, as is usually required for
models of elliptical galaxies.
This mass distribution is described by a central velocity dispersion $\sigma_0$ 
related to $\rho_0$ for a circular potential by the following relation:
\begin{equation}
\label{rhodesigma}
\rho_0=\frac{\sigma_0^2}{2\pi\mathrm{G}}\left(\frac{r_{\mathrm{cut}}+r_{\mathrm{core}}}{r_{\mathrm{core}}^2 r_{\mathrm{cut}}}\right)
\end{equation}
It is easy to show that for a vanishing core radius, the
density profile obtained becomes identical to the density
profile used by \textsc{bbs} for modeling galaxy-galaxy lensing. Since many
authors are using the same mass profile used by \textsc{bbs} in their
galaxy-galaxy lensing studies, it allows for easy comparison with our
results.

\subsection{Cluster description}

We describe the cluster component as a large scale smooth
component (see Paper I for details).  This cluster component is put in by
hand and we use the results found by \citet{sebphd}
on the same data set as parameters. 

\subsection{Maximum Likelihood Analysis}

The details of the method have been given in Paper I.  
Here we give a brief outline of the method.

Once we have the image catalogue, we process it through a numerical code
that retrieves the input parameters of the lenses using a maximum
likelihood method as proposed by \citet{rix}.  For each image ($i$),
given a mass model for the foreground lenses galaxies (\emph{e.g.}
$\sigma_0$, $r$), we can evaluate the amplification matrix $a_i$ as a
contribution of all the foreground galaxies $j$ ; $z_j<z_i$ that lies
within a circle of inner radius $R_{\mathrm{min}}$, and outer radius
$R_{\mathrm{max}}$ and of centre the position of the image ($i$):
\begin{equation}
a_i(\sigma_0,r)=\sum_{\begin{array}{c} {z_j<z_i} \\ d(i,j)<R_{\mathrm{max}}\end{array}} a_{ij}(\sigma_0,r)
\end{equation}
Given the observed ellipticity $\vec{\varepsilon^i_{\mathrm{obs}}}$ and the associated amplification
matrix $a_i$, we are able to retrieve the intrinsic ellipticity $\vec{\varepsilon_i^s}$
of the source before lensing:
\begin{equation}
\vec{\varepsilon_i^s}=F\left(\vec{\varepsilon^i_{\mathrm{obs}}}, a_i(\sigma_0,r)\right)
=\vec{\varepsilon_i^s}(\sigma_0,r)
\end{equation}
In order to assign a likelihood to the parameters used to describe the lense
galaxies, we use
$P^s$, the ellipticity probability distribution in the absence of
lensing.  Doing that for each image of the catalogue, we construct the
likelihood function:
\begin{equation}
\label{L}
\mathcal{L}(\sigma_0,r)=\prod_i P^s(\vec{\varepsilon^s_i})
\end{equation}
which is a function of the parameters used to define the mass models
of the lenses. For each pair of parameters, we can compute a
likelihood.
The larger this function, the more likely the parameters used to describe
the lenses.  See Paper I for a discussion on the convergence 
properties of this likelihood function.

As exposed in detail in Paper I, we found that galaxy-galaxy lensing studies
are first sensitive to the mass enclosed within a given radius; this gave us the idea
to re-parameterize the problem in terms of more physical quantities: the aperture mass
calculated in an aperture radius. Instead of fitting the deformations
in the ($\sigma_0, r_{\mathrm{cut}}$) plane, we can 
fit them directly in 
the ($M_{\mathrm{aper}}, R_{\mathrm{aper}}$) plane.
We have:
\begin{equation}
\label{ML_S_R}
\mathcal{L}=\mathcal{L}(\sigma_0, r_{\mathrm{cut}})\quad \& \\ 
M_{\mathrm{aper}}=M_{\mathrm{aper}}(R_{\mathrm{aper}},\sigma_0,r_{\mathrm{cut}})
\end{equation}
so we can write:
\begin{equation}
\sigma_0=\sigma_0(M_{\mathrm{aper}},R_{\mathrm{aper}},r_{\mathrm{cut}})
\end{equation}
the likelihood function then becomes:
\begin{equation}
\mathcal{L}(M_{\mathrm{aper}},R_{\mathrm{aper}},r_{\mathrm{cut}})
\end{equation}
and by summing over $r_{\mathrm{cut}}$, we obtain:
\begin{equation}
\mathcal{L'}=\sum_{r_{\mathrm{cut}}} \mathcal{L}(M_{\mathrm{aper}},R_{\mathrm{aper}},r_{\mathrm{cut}})
\end{equation}
this means that:
\begin{equation}
\mathcal{L'}=\mathcal{L'}(M_{\mathrm{aper}},R_{\mathrm{aper}})
\end{equation}
The new likelihood function is then a function of the aperture mass and the aperture
radius. Note that $M_{\mathrm{aper}}$ and $R_{\mathrm{aper}}$ are not independent
parameters, which explains the shape of the likelihood contours in the 
($M_{\mathrm{aper}}, R_{\mathrm{aper}}$) plane: they remain open along the
$R_{\mathrm{aper}}$ axis.

In Paper I we tested extensively the maximum likelihood method on
simulated data defined to match observations to study the accuracy
with which input parameters for mass distributions for galaxies can be
extracted. We showed that the two standard parameters that
characterise galaxy halo models, the central velocity dispersion and
the truncation radius can be retrieved reliably from the maximum
likelihood analysis and that the proposed re-parameterization allows us
to put strong constraints on the aperture mass of a galaxy halo (with
less than 10\% error). Thus we are confident in applying this method
to the data set presented in Section 3 and 4.

\section{Results of the galaxy-galaxy lensing study}

In this Section, we present the results we have for the
elliptical galaxies in the field of the different clusters studied in this paper.
In the following plots, the likelihood contours are the 1$\sigma$, 2$\sigma$
and 3$\sigma$ confidence level contours.

\subsection{Null tests}
In order to examine the validity of the galaxy-galaxy lensing signal,
we generated a non physical catalogue as follows:
\begin{itemize}
\item the orientation angle $\theta$ for the galaxies is randomly assigned
\item the position of galaxies that verify $z_{\mathrm{bayes}}>0.5$ is randomly assigned
\item the position of the foreground cluster elliptical galaxies is
randomly assigned.
\end{itemize}
In each case, the likelihood function does not exhibit any significant maximum.

\subsection{Catalogues}
The catalogues used in the maximum likelihood analysis are formed by the 
objects with the following characteristics:
\begin{itemize}
\item 3 colored objects (it means objects detected in the three bands: \textsc{b}, \textsc{r}, \textsc{i})
\item shape parameters coming from the \textsc{r} band which appeared to 
be the less noisy one
\item errors on the estimation of the ellipticity lower than 0.1
\item redshift assigned at the redshift of the cluster for ellipticals galaxies, 
Bayesian photometric 
redshifts for the background population ($z>0.5$).
\end{itemize}
The average number of lenses in the considered catalogues is $\sim$ 700,
and the average number of background sources is $\sim$ 7500.

This catalogue is the input to the maximum likelihood code.
We use the scaling relations given in Section 2.2, with
$\delta=0.25$ and $\alpha=0.5$.
From theoretical considerations, the $R_{\mathrm{max}}$ parameter should 
be of the order of 100$\arcsec$ (see Paper I).

\subsection{Results}

\begin{figure}[h!]
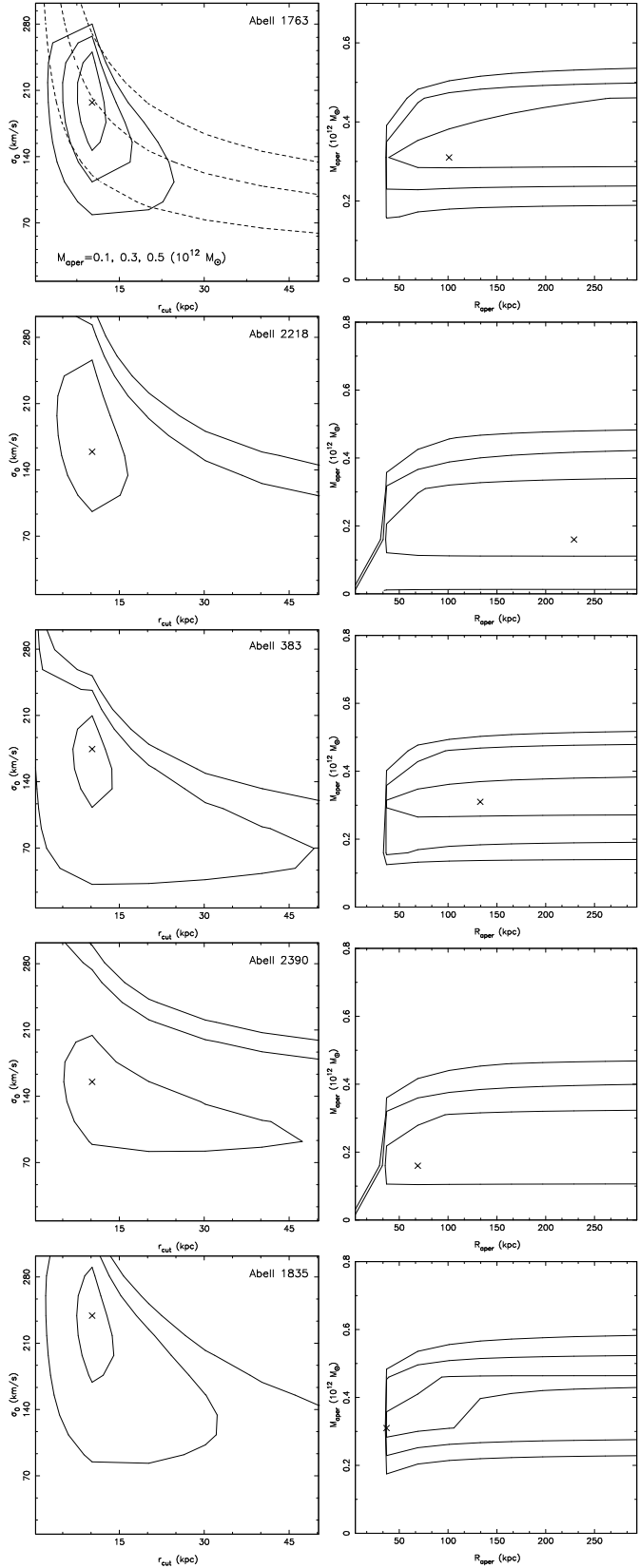

\begin{center}
\includegraphics[height=4.35cm,width=4.35cm]{5543f5.1.ps}
\includegraphics[height=4.35cm,width=4.35cm]{5543f5.2.ps}\\
\includegraphics[height=4.35cm,width=4.35cm]{5543f6.1.ps}
\includegraphics[height=4.35cm,width=4.35cm]{5543f7.1.ps}\\
\includegraphics[height=4.35cm,width=4.35cm]{5543f6.2.ps}
\includegraphics[height=4.35cm,width=4.35cm]{5543f7.2.ps}\\
\includegraphics[height=4.35cm,width=4.35cm]{5543f6.3.ps}
\includegraphics[height=4.35cm,width=4.35cm]{5543f7.3.ps}\\
\includegraphics[height=4.35cm,width=4.35cm]{5543f6.4.ps}
\includegraphics[height=4.35cm,width=4.35cm]{5543f7.4.ps}\\
\end{center}
\caption{Results of the galaxy-galaxy lensing analysis in the ($\sigma_0$, $r_{\mathrm{cut}}$) 
plane (left) and in the ($M_{\mathrm{aper}}$, $R_{\mathrm{aper}}$) plane. From top to bottom,
Abell~1763, Abell~2218, Abell~383, Abell~2390, Abell~1835. In the case of Abell~1763, 
along the dotted lines in the ($\sigma_0, r_{\mathrm{cut}}$) plane, the aperture mass computed in an
aperture radius of 100 kpc is kept constant equal to the value indicated on the plot}
\label{results}
\end{figure}

Results are presented in Fig.~\ref{results}. We fit the
deformations with a \textsc{piemd} profile, in the ($\sigma_0,
r_{\mathrm{cut}}$) plane and in the
($M_{\mathrm{aper}}, R_{\mathrm{aper}}$) plane. 
Table~\ref{table1} summarizes the
results we obtained.  
The main results are the
following: (i) we fit reasonable values for the velocity dispersions,
around 200 km\,s$^{-1}$. This is reasonable in the sense that it is
comparable to values inferred using more traditional methods (rotation
curves, \textsc{x}-ray observations, satellite dynamics) (ii) we find dark
matter halos to be very compact compared to field galaxies of
equivalent luminosity: considering all cluster galaxies halos, an
upper limit on the truncation radius is set at 50 kpc (\textsc{piemd} results
on Abell~383), when the truncation radius inferred on field galaxies is
found to be larger than a few hundreds of kpc (see Section 2 and
Fig.~\ref{comparfieldcluster}).
As mentioned in Section 2.2, the truncation radius is related to the extension
of the halo. We can say that cluster galaxies are more compact than field 
galaxies because the slope of their mass profile steepens earlier, thus the
corresponding mass profile reaches a low density value earlier.
As a consequence, dark matter halo of cluster galaxies are less extended than
they are in the case of field galaxies.

Galaxy cluster Abell~1689 was also studied as part of the galaxy cluster
sample. We found similar constraints for galaxy halos living in this cluster,
but the significance of the detection is below the $1\sigma$ level, so we do not 
add this detection in this paper.

It should also be pointed out that detections for the different cluster
galaxy dark matter halos are comparable to one another; this is due to
the fact that these different clusters all have similar physical
properties and constitute a very homogeneous sample.

\begin{table*}
\centering{
\begin{tabular}[h!]{ccccc}
\hline
\hline
\noalign{\smallskip}
Cluster & $\sigma_0^*$, {\footnotesize km\,s$^{-1}$ (\textsc{piemd})} & $r_{\mathrm{cut}}^*$, {\footnotesize kpc (\textsc{piemd})} & {\footnotesize (M/L)$^*$} \\
\noalign{\smallskip}
\hline
\hline
\noalign{\smallskip}
\noalign{\smallskip}
A1763 & 200$^{+70}_{-115}$ (3$\sigma$) & $\leq$ 25 (3$\sigma$) & 19$^{+16}_{-6}$ (3$\sigma$) \\
\noalign{\smallskip}
\noalign{\smallskip}
\hline
\noalign{\smallskip}
\noalign{\smallskip}
A1835 & 240 $^{+81}_{-159}$ (2$\sigma$) & $\leq$ 32 (2$\sigma$) & 20$^{+18}_{-6}$ (3$\sigma$) \\
\noalign{\smallskip}
\noalign{\smallskip}
\hline
\noalign{\smallskip}
\noalign{\smallskip}
A2218 & 200$^{+96}_{-64}$ (1$\sigma$)& $\leq$ 18 (1$\sigma$) &  13$^{+10}_{-12}$ (2$\sigma$) \\
\noalign{\smallskip}
\noalign{\smallskip}
\hline
\noalign{\smallskip}
\noalign{\smallskip}
A383 & 175 $^{+66}_{-143}$ (2$\sigma$) & $\leq$ 50 (2$\sigma$) & 20$^{+13}_{-10}$ (3$\sigma$) \\
\noalign{\smallskip}
\noalign{\smallskip}
\hline
\noalign{\smallskip}
\noalign{\smallskip}
A2390 & 155$^{+50}_{-75}$ (1$\sigma$) & $\leq$ 47 (1$\sigma$) & 10$^{+21}_{-4}$ (1$\sigma$)\\
\noalign{\smallskip}
\noalign{\smallskip}
\hline
\hline
\end{tabular}
\caption{Summary of the detections, for a $L^*$ luminosity. The mass corresponds to the total mass
computed with a \textsc{piemd} profile, and luminosity comes from the \textsc{r} band. Here $\sigma$ corresponds to the confidence level of the detection.}
\label{table1}
}
\end{table*}

\subsection{Comparison with other results on cluster galaxies}
\begin{figure}[h!] 
\begin{center} 
\includegraphics[width=8cm,height=8cm]{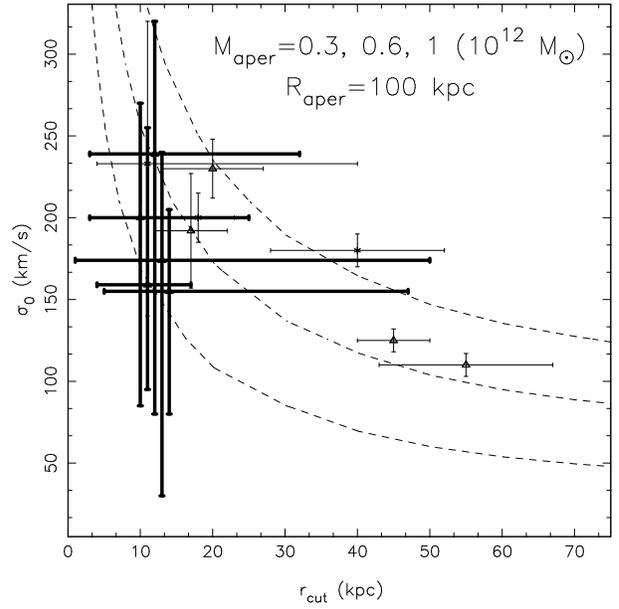}
\end{center}
\caption{Comparison of our results (black) with the results
from Natarajan et al. and Geiger \& Schneider (grey). Note that a comparison is possible keeping
in mind that the different data sets and the methods used are different}
\label{comparit}
\end{figure}

Fig.~\ref{comparit} shows 
a comparison of our results (in black) with results 
on cluster galaxies from Natarajan et al. and from
Geiger \& Schneider (in grey).
There is a good agreement between the
different studies, though the data sets as well as the analysis are
quite different: 
as mentioned before studies by
Natarajan et al. and Geiger \& Schneider are based on \textsc{hst} data and
therefore they probe central cluster galaxies.
Natarajan et al. used the constraints derived from the
observations of multiple images, whereas Geiger \& Schneider
did not include strong lensing constraints in their analysis.
Our results are averaged on a cluster galaxy population from the centre to 
$\sim$ 2 Mpc, thus we probe the whole centre of the cluster as well as part of the
transition region.

The results presented in this work can also can be compared to the
constraints found by \citet{smith05} on the cluster galaxies
of a galaxy cluster sample that contain the galaxy clusters we study
in this work. \citet{smith05} presented an 
analysis of 10 \textsc{x}-ray luminous galaxy clusters based on \textsc{hst} observations.
Therefore they probe the inner part of the galaxy cluster. From the
observation of multiply imaged systems, they modeled the mass
distribution in the cluster cores ($R<500$ kpc). In order to reproduce
the location of multiple images detected on the image, each cluster
model comprised of a number of parameterized mass components which account for 
mass distributed on both cluster and galaxy scales.
Therefore, their study also gives some constraints on the characteristic
parameters of the cluster galaxies. Using a \textsc{piemd} profile, they found:
$\sigma_0 = 180 \pm20$ km\,s$^{-1}$ and $r_{\mathrm{cut}}=23$ kpc, in very
good agreement with the results presented here.
 
The strong lensing modeling of galaxy cluster Abell~1689 from \textsc{hst acs} 
observations by \citet{halkola} also provides evidence for cluster galaxies to be 
significantly stripped, with a cut radius of 24 kpc.

The work by \citet{rachel06} also gives some insight on the extension of the dark matter
halos:
for early type galaxies living in high density regions,
they probe the extension of the dark matter halos by searching for a depression
in the lensing signal relative to that for field galaxies on 50-100 $h^{-1}$
kpc scales. A large amount of tidal stripping in clusters would cause a
depression in the lensing signal on scales below the virial radius.
No such depression was found by the authors, ruling out scenarios that have
most satellites strongly stripped.
This result seems to be in disagreement with the one presented in this paper and with early
results of galaxy-galaxy lensing in clusters.
It should be noted that in order to look for a depression in the lensing signal, one needs
to use a direct averaging method, which is not well suited
for studying cluster galaxies that are very close one to each other 
compared to field galaxies (see Section 2.1).
Moreover, as mentioned before, a dark matter halo with $r_{\mathrm{cut}}\sim50$ kpc
(as found in this work)
still has a significant amount of mass below 50\,kpc and will generate
a non negligible lensing signal below 50\,kpc.
Moreover, it is worth mentioning
the fact that we do present some constraints for galaxies inhabiting very massive
clusters where tidal stripping is expected to be more efficient, thus our results
are biased towards very high density environments.
Therefore the different results are not easy to compare and may not be in
disagreement.

\section{Discussion \& Conclusion}

The results presented here come from wide field ground based data.
As a consequence, we probe the cluster galaxy population
out to a larger radius than previous studies 
($\sim 2$ Mpc).
This means that our results are averaged over a galaxy population
located all the way from the centre of the cluster to the transition region of the cluster.
Our study confirms the fact that galaxy halos in clusters are
significantly more compact compared to halos around field galaxies of equivalent luminosity.

The results presented in this work are in very good agreement
with numerical simulations.
A detailed comparison with realistic numerical simulations 
will be presented in a forthcoming publication
(see also recent work by Natarajan et~al., 2006).
The theoretical expectation is that the global tidal field of a
massive, dense cluster potential well should be strong enough to
truncate the dark matter halos of galaxies that traverse the cluster
core. As mentioned before, this expectation has been observationally
confirmed by several independent previous studies. Early numerical work (see
\emph{e.g} \citealt{merritt83}, \citealt{richstone84}) noticed that a large fraction of the
mass initially attached to galaxies in the central megaparsec is
stripped.  
\citet{bullock} found that halos in dense environments are more truncated than their
isolated counterparts of the same virial mass.
Avila-Reese et~al. (1999, 2005) found that halos in cluster regions are more
concentrated than isolated halos.
\textsc{n}-body simulations of cluster formation and evolution
(\citealt{ghigna98}, \citealt{ghigna00}) find that the dominant interactions are between the
global cluster tidal field and individual galaxies after $z=2$. The
cluster tidal field significantly strips galaxy halos.
As previously noted, we probe in this paper lens galaxies located all the way from the centre of the
cluster to $\sim$ 2 Mpc. This galaxy population is bound to the cluster and it is reasonable
to think that it has experienced the cluster potential at least once:
the characteristic time for a galaxy to cross a cluster is about $10^{9}$ years, and the age of a galaxy 
cluster that formed at $z=1$ is about $10^{10}$ years.

To conclude, we presented the first galaxy-galaxy lensing results to
date that probe cluster galaxies from a ground based survey.  This
study has confirmed the fact that galaxy halos in clusters are
significantly less massive and more compact compared to galaxy halos
around field galaxies of equivalent luminosity. Moreover, this
confirmation is based on the analysis of 5 massive clusters lenses
whose properties are close to each other, hence the confirmation
we provide is a strong one since it relies on a homogeneous sample.

\begin{acknowledgements}
The Dark Cosmology Centre is funded by the Danish National Research
Foundation. ML wishes to acknowledge Roser Pell\`o for help in
using \textsc{hyperz}, in understanding photometric redshift
determination and for useful discussions and encouragements. ML 
also thanks many people for carefully reading this paper and
for constructive comments, in particular: Alexie Leauthaud, 
\'Ard\'is El\'iasd\'ottir, 
Gary Mamon, Jens Hjorth and Jesper Sommer-Larsen.
PN acknowledges gratefully support from \textsc{nasa} via \textsc{hst} grant HST-GO-09722.06-A.
Argelander-Institut f\"ur Astronomie is founded by merging of the Institut f\"ur Astrophysik
und Extraterrestrische Forschung, the Sternwarte, and the
Radioastronomisches Institut der Universit\"at Bonn.
IRS and GPS acknowledge support from the Royal Society.
\end{acknowledgements}

\renewcommand{\refname}{References} \bibliographystyle{aa}

\label{lastpage}

\end{document}